\documentclass[runningheads]{llncs}
\usepackage[T1]{fontenc}
\usepackage{graphicx}
\usepackage{booktabs} 
\usepackage{longtable}
\usepackage{comment}
\usepackage{todonotes}
\begin{document}
%
%Towards contextualized and multifaceted code review at Ericsson
%\title{Towards contextualized and multifaceted code review at Ericsson}
\title{Using Agentic AI for contextualized and multifaceted code review at Ericsson}
\titlerunning{Agentic AI for contextualized and multifaceted code review}
\author{Muhammad Laiq\inst{1} \and
Ricardo Britto\inst{1,2} \and
Muhammad Usman\inst{1} \and
Nishrith Saini \inst{2} \and
Deepika Badampudi \inst{1}}
\authorrunning{Laiq et al.}
\institute{Blekinge Institute of Technology, Sweden. \\
\email{\{muhammad.laiq, muhammad.usman, ricardo.britto, deepika.badampudi\}@bth.se} \and
Ericsson AB, Sweden. \\
\email{\{ricardo.britto, nishrith.saini\}@ericsson.com}}
\maketitle  % typeset the header of the contribution

\begin{abstract}
\textbf{Context:} Conducting effective code reviews is increasingly challenging due to the growing complexity of software systems and the accelerated code generation by AI coding agents. LLM-based approaches for code reviews have shown promising results in identifying defects and improving code quality. However, existing approaches rarely consider project-specific contextualized knowledge, and few have been evaluated in industrial settings.  %Although there is extensive research on supporting developers in code review, many studies provide only narrow insights. In addition, the use of context-specific knowledge and the evaluation of the proposed solutions in real industrial settings remain limited.

\textbf{Objective:} In this study, we propose a multi-agent–based solution that provides multifaceted assessments of code changes. 

\textbf{Method:} Following the Design Science Research Process, we developed and evaluated our solution in an industrial setting. Our solution combines specialized agent skills with context-specific knowledge to identify antipatterns in code changes across four dimensions: readability, maintainability, reliability, and performance. Using our solution, we generated reviews for several code commits and identified more than 200 issues. These issues were then manually validated by the developers of the case company for their correctness and importance. 

\textbf{Results:} The evaluation results show that our solution achieves 96\% accuracy in correctly identifying issues in the investigated code commits. Furthermore, around 69\% of the correctly identified issues were rated as important, with approximately 33\% rated as severe issues that must be fixed and 36\% as important issues that should be fixed. Qualitative feedback from developers corroborates these findings and highlights the usefulness of the generated reviews.

\textbf{Conclusion:} Our findings provide empirical evidence from an industrial evaluation that combining specialized agent skills with context-specific knowledge yields accurate, practically useful code reviews.

% Old abstract
% Although there is extensive research on supporting developers in code review, many studies provide only narrow insights. In addition, the use of context-specific knowledge and the evaluation of the proposed solutions in real industrial settings remain limited. In this study, we propose a multi-agent–based solution that provides multifaceted assessments of code changes.  Our solution combines specialized agent skills with context-specific knowledge to identify antipatterns in code changes across four dimensions: readability, maintainability, reliability, and performance. We develop and evaluate our solution in an industrial setting. Using it, we generated reviews for several code commits of multiple projects and identified more than 200 issues. These issues were then manually validated by the developers of the case company for their correctness and importance. The evaluation results show that our solution achieves 96\% accuracy in correctly identifying issues in the investigated code commits. Furthermore, around 69\% of the correctly identified issues were rated as important, with approximately 33\% rated as severe issues that must be fixed and 36\% as important issues that should be fixed. Qualitative feedback from developers corroborates these findings and highlights the usefulness of the generated reviews. Our findings provide empirical evidence from an industrial validation that combining specialized agent skills with context-specific knowledge yields accurate, practically useful code reviews.

\end{abstract}

\section{Introduction}
Code review is a fundamental practice in software engineering that helps improve software quality and facilitates knowledge sharing among developers \cite{bavota2015four,bosu2016process,badampudi2023modern}. Developers routinely perform code reviews to assess code changes, for example, to identify defects, maintain coding standards, and ensure the long-term health of software systems. However, conducting effective reviews remains a challenge. Reviewers are expected to understand the intent of code changes, reason about their impact on the broader system, and identify potential issues, all while working within fast-paced development cycles. This imposes a significant cognitive load on the reviewers. The challenge becomes even greater as software systems continue to grow in scale and complexity, making it increasingly difficult to maintain thorough, consistent, and timely reviews \cite{kononenko2016code,bacchelli2013expectations,kudrjavets2022mining}.
This challenge is further amplified by the rise of agentic AI in software development. As organizations delegate an increasing share of implementation work to AI coding agents, code is produced at a speed and volume that human reviewers cannot match, making reviewers a bottleneck and increasing their cognitive load. This is particularly acute in companies moving toward a fully AI-native software development approach, where orchestrated agent workforces operate in parallel, and assurance is expected to keep up with code generation rather than lag behind \cite{britto2026ai}. %The challenge is even greater in InnerSource projects, where core teams evaluate contributions from developers who are not part of the project. While the reviewers have knowledge of the project, they must ensure that external contributions align with the project's specific conventions, architecture, and design decisions. This situation heightens the need for review support that utilizes this contextual knowledge effectively.

To assist practitioners in code review, several automated solutions have been proposed in the literature \cite{yang2026roadmap,badampudi2023modern}. These solutions include static analysis tools, traditional machine learning and deep learning-based approaches, and more advanced large language models (LLMs). Although prior work has made significant progress and shown positive results, important gaps remain. For example, many of these approaches focused on providing a narrow review \cite{unterkalmsteiner2024help,yang2026roadmap}. In practice, reviewers rarely assess code changes from a single perspective. Instead, they apply a multifaceted reasoning process that considers a range of software quality concerns simultaneously. For example, maintainability, readability, and other non-functional qualities. Another important challenge concerns the role of contextual knowledge in review activities. Human reviewers rely extensively on contextual information, including system architecture, organizational standards, dependencies, and domain-specific constraints, when assessing the implications of a code change. Previous work also emphasizes that reviewers need both the local and global context to effectively understand and assess changes \cite{unterkalmsteiner2024help}. However, much of the current work uses limited context-specific knowledge \cite{bacchelli2013expectations,yang2026roadmap,unterkalmsteiner2024help}. As a result, many existing approaches struggle to provide feedback that reflects the broader context in which software systems evolve. In addition, evaluations of the proposed approaches in real industrial settings remain relatively limited.

This study aims to contribute towards filling the above-mentioned gaps. We propose a contextualized multi-agent code review approach. The proposed approach combines multiple specialized agents with context-specific knowledge to support a multifaceted assessment of code changes. Within the scope of this study, we assess code changes across four dimensions: readability, maintainability, reliability, and performance. The study follows the Design Science Research Process by Offermann et al. \cite{offermann2009outline}. The solution is developed and evaluated in an industrial context (Ericsson).

In this study, we answered the following research questions.

\begin{itemize}
    \item \textbf{RQ1:} What design choices are essential for providing multifaceted and contextualized code review?
    \item \textbf{RQ2:} How accurate is the proposed approach in identifying issues from code commits?
    \item \textbf{RQ3:} How effective is the proposed approach in terms of identifying important issues from code commits?
    
\end{itemize}

We assess accuracy by measuring the proposed approach's ability to identify valid issues in code changes, with each identified issue labeled correct or incorrect, and effectiveness by measuring the extent to which correctly identified issues are considered worth fixing, with each correct issue rated as minor, medium, or high.

This paper is organized as follows. Section \ref{sec:relatedwork} describes the related work on the topic. Sections \ref{sec:method}--\ref{sec:solution} present the research design and the proposed solution. Section \ref{sec:results} presents the results of the study. Section \ref{sec:discussion} discusses the findings of the study. Section \ref{sec:validitythreats} describes the threats to validity. Finally, Section \ref{sec:conclusion} concludes the paper with future work.

\section{Related work}\label{sec:relatedwork}
There is a plethora of research on code review that spans several dimensions, including automated reviewer assignment, code comment analysis, and code change assessment \cite{yang2026roadmap,badampudi2023modern}. Among these areas, the most relevant work for this study is the assessment of code changes to provide review comments.

Significant research has focused on assessing code changes to generate code review comments. Early automated code review approaches primarily relied on static analysis and rule-based techniques. Such approaches focused on identifying predefined patterns, coding standard violations, or structural issues in source code \cite{johnson2013don,beller2016analyzing}. Although these tools are effective for detecting specific categories of issues, they generally provide limited support for higher-level reasoning about software quality. In practice, these approaches often operate independently of the broader development and architectural context in which code changes occur. Subsequent research explored machine learning and deep learning approaches to automate code review. For example, Gupta et al. \cite{gupta2018intelligent}, Li et al. \cite{li2019deepreview}, and Shi et al. \cite{shi2019automatic} used deep learning-based approaches to automate code review. In particular, Convolutional Neural Networks and Long Short-Term Memory were used to build models. Although these approaches have improved automated code review compared to previous methods, their capabilities are largely limited by the quality and scope of training data, and they often struggle to generalize beyond localized patterns.

Recent advances in LLMs have substantially changed the landscape of automated code review research. Due to their ability to process source code and natural language jointly, LLM-based approaches have shown improved performance in code understanding, review comment generation, and reasoning about code changes \cite{fan2025exploring,ramesh2025automated,cihan2025automated,yang2026roadmap,ren2025hydra,sun2025bitsai}. For using LLMs in this task, different strategies have been investigated in the literature. For example, Nashaat and Miller \cite{nashaat2024towards} fine-tuned pretrained models such as T5 and CodeT5 for code review, while Ramesh et al. \cite{ramesh2025automated} investigated prompt-based approaches using two variants of Llama models without task-specific fine-tuning. Other studies proposed multi-agent and tool-augmented architectures to improve review reasoning \cite{ren2025hydra,sun2025bitsai}.

Despite significant progress and promising results from LLM-based approaches to code review, several gaps still exist. For example, a recent survey by Yang et al. \cite{yang2026roadmap} found that current approaches often use limited contextual information. Many rely mainly on isolated diffs or local code snippets, while effective reviews often depend on organizational and repository-specific context. Furthermore, they noted that much of the work provides only a narrow review. Only a few studies, such as those by Ren et al. \cite{ren2025hydra} and Sun et al. \cite{sun2025bitsai}, have moved toward a multidimensional review through taxonomies. Similar gaps have also been reported by Unterkalmsteiner et al. \cite{unterkalmsteiner2024help}. Finally, we observe that industrial evaluations of LLM-based code review approaches remain relatively limited. Much of the current literature relies on offline experiments or open-source datasets.

Consequently, this study proposes a multifaceted contextualized code review approach. The proposed approach uses multiple specialized agents, each targeted at different aspects of code changes. For example, an agent for assessing code maintainability and an agent for assessing code reliability. In addition, the approach integrates system-specific knowledge via a code knowledge graph and reusable agent skills, enabling agents to reason beyond submitted code diffs. The proposed solution is developed and evaluated in the real industrial setting, providing evidence of its practical applicability beyond offline experiments.

\section{Research design}\label{sec:method}
%We follow the design science methodology proposed by Wieringa...

This study employs the Design Science Research Process (DSRP) proposed by Offermann et al. \cite{offermann2009outline} to develop and evaluate technological artifacts. We operationalize DSRP through three phases: problem identification, solution design and development, and evaluation (Sections \ref{sub:phase1}--\ref{sec:userstudy}). The solution (resulting artifact) is described in detail in Section \ref{sec:solution}, and the results of the evaluation are reported in Section \ref{sec:results}.

\subsection{Phase 1: Problem identification}\label{sub:phase1}
In this phase, the problem was identified by reviewing the relevant literature (see details in Section \ref{sec:relatedwork}). Beyond the literature, the problem was discussed with practitioners at the case company to understand current review practices and the challenges reviewers face, particularly as AI coding agents produce an increasing share of code. Three practitioners were involved: one manager and two senior developers. These discussions confirmed the problem's practical relevance and, together with the literature, informed the solution's scope. For this initial iteration, it was decided to focus on the four dimensions of the code review and to ground the assessment of each dimension in an established antipattern catalog (see details in Section \ref{sub:Review_dimensions}). %Following the evaluation, it was decided to extend the solution in the next iteration with additional quality aspects, such as security and technical debt; a feedback mechanism that learns from developers' assessments; and support for InnerSource projects, where contributions from outside the core team must align with the project's conventions and architecture. Evaluating the solution in other languages and projects was also identified as necessary.

\subsection{Phase 2: Solution design and development}\label{sub:phase2}
This phase translates the identified problem into a working artifact: a contextualized multi-agent solution that provides a multifaceted assessment of code changes. The artifact is based on two main design decisions. First, the review is decomposed into specialized agents, each dedicated to one quality dimension and guided by its own prompt, antipattern catalog, and review instructions, which addresses the narrow-review gap. Second, the agents do not reason about isolated diffs. A context builder supplies each agent with a contextual view constructed from the code repository and a code knowledge graph that addresses the limited-context gap. The artifact, its components, and the antipattern catalogs used by each agent are described in Section \ref{sec:solution}.

\subsection{Phase 3: Evaluation}\label{sec:userstudy}
The evaluation was designed as a static validation \cite{gorschek2006model}, in which the solution is applied in a realistic industrial context (Ericsson). We generated reviews for seven code commits from four different projects at the case company. Two of these projects were small, and two were large. In total, 206 issues were identified from the selected code commits. We then conducted a user study to evaluate their correctness (RQ2) and importance (RQ3). The issues were reviewed by the developers of the code commits, all of whom are senior developers. We used the template shown in Table \ref{tab:feedback-template} to collect feedback from the developers on issues identified. In total, five developers reviewed the 206 issues: four of them reviewed one commit each, with 36, 36, 27, and 11 issues, respectively, and one reviewed three commits with 96 issues in total.

\begin{table}[!ht]
    \centering
    \scriptsize
    \caption{Feedback template for user study}
    \vspace{-2mm}
    \label{tab:feedback-template}
    \begin{tabular}{l}
    \toprule
\textbf{\underline{Part I: Details of an identified issue}} \\
\textbf{Antipattern:} Antipattern/Name of the identified issue, e.g., The Blob. \\
\textbf{Location:} Location of the identified issue, e.g., file name and line numbers.  \\
\textbf{Description:} Detailed description of the issue.  \\
\textbf{Fix suggestion:} Fix suggestion. \\ \\

\textbf{\underline{Part II: User feedback for the identified issue}} \\
\textbf{A. Correctness of the identified issue} \\
--- Correct [ ] \\
--- Incorrect [ ] \\ 

\textbf{B. Importance of the identified issue}  \\
--- \textbf{Minor:} Minor or low-impact suggestion [ ] \\
--- \textbf{Medium:} Important issue that should be fixed [ ] \\
--- \textbf{High:} Severe issue that must be fixed [ ] \\
\textbf{C. Free text field:} Optional feedback  \\ 
\bottomrule
	
	\end{tabular}
\end{table}

\section{The proposed solution}\label{sec:solution}
\begin{figure}[!ht]
    \centering
    \includegraphics[width=0.95\textwidth]{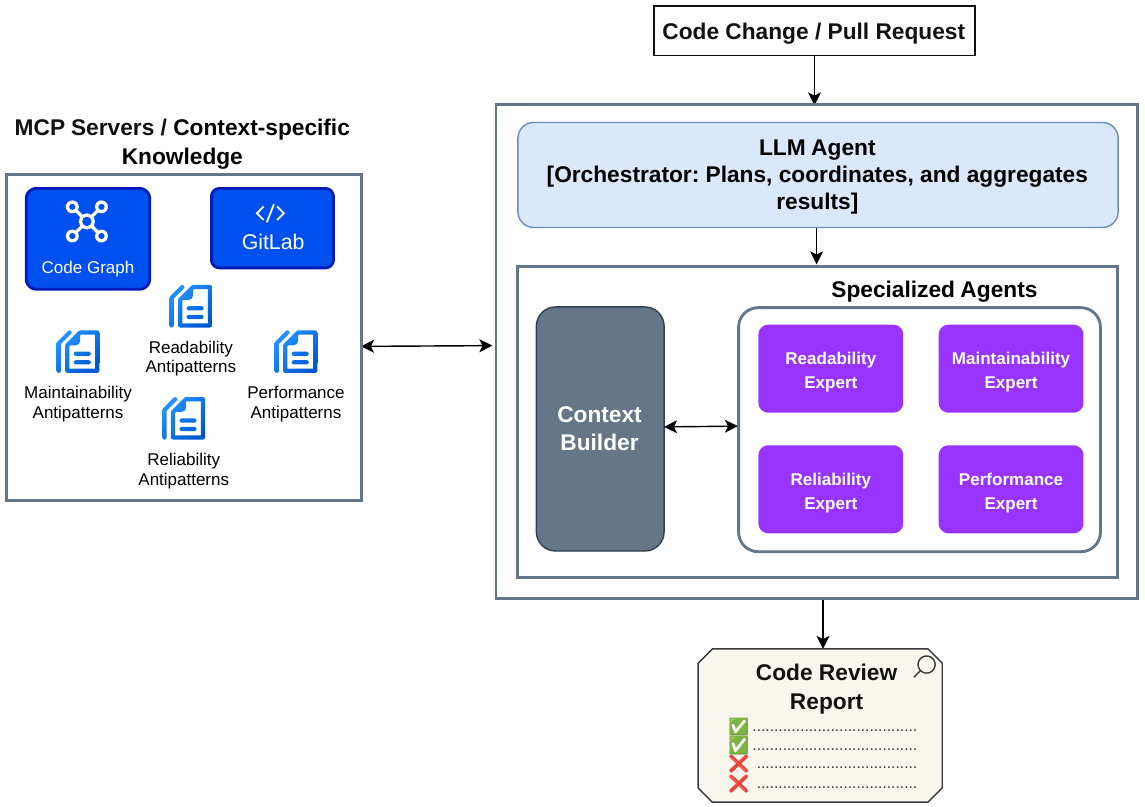}
    \vspace{-2mm}
    \caption{Overview of the proposed framework}
    \label{fig:core}
\end{figure}

Figure \ref{fig:core} presents an overview of the proposed framework. The framework aims to provide a multifaceted assessment of code changes using specialized code review agents. Each agent assesses code changes considering repository-specific contextual information. Although the framework can be readily extended to additional code quality aspects, this study focuses on the following four key dimensions, each one with a dedicated AI agent: readability, maintainability, reliability, and performance.

\subsection{Code review dimensions focused}
\label{sub:Review_dimensions}
\paragraph{\textbf{A. Readability Expert:}} The readability agent is focused on assessing whether the modified code is easy to understand and follow. The assessment focuses on several aspects, such as naming clarity, structural organization, nesting complexity, code formatting, visual readability, and the use of understandable programming constructs. To guide the analysis, the agent relies on the readability antipatterns summarized in Table \ref{tab:antipatterns-readability}. These antipatterns have previously been used in the literature \cite{sergeyuk2024reassessing,sergeyuk2024assessing} to assess code readability.

\begin{table}[!ht]
    \centering
    \scriptsize
    \caption{Antipatterns for readability \cite{sergeyuk2024reassessing,sergeyuk2024assessing}}
    \vspace{-2mm}
    \label{tab:antipatterns-readability}
    \begin{tabular}{p{11.9cm}}
    \toprule
\textbf{Code structure:} Logical separation of functionality vs. tangled code. \\
\textbf{Nesting:} Flat, linear code vs. deeply nested blocks.  \\
\textbf{Understandable Goal:} Clear task or function vs. ambiguous purpose.  \\ 
\textbf{Code Length:} Concise and readable vs. unnecessarily long code.  \\
\textbf{Inline Actions:} One action per line vs. multiple actions in the same line.  \\
\textbf{Reading Flow:} Code reads well from top to bottom, or while reading, the eyes jump from top to bottom and back up again.  \\
\textbf{Sufficient Contextual Info:} Code is not sufficiently explained and needs additional info to understand what it does, or code is overexplained.  \\
\textbf{Code Style:} Code conforms to style guides, or code is poorly formatted. \\
\textbf{Magic Numbers:} Code uses named constants or code uses magic numbers.  \\
\textbf{Naming:} Naming clarifies code functionality, or naming is confusing. \\
\textbf{Code Patterns:} Code uses basic, known code patterns, or code looks unfamiliar or nonstandard. \\ 
\textbf{Visual Organization:} There is balance in the color blocks, or there are huge chunks of color blocks that stand out in a distracting way. \\ 
\bottomrule
	
	\end{tabular}
\end{table}

\noindent\paragraph{\textbf{B. Maintainability Expert:}} The maintainability agent focuses on long-term software evolution concerns. In particular, it identifies issues that may increase future maintenance effort, reduce modularity, or complicate code evolution. The agent uses classical maintainability antipatterns \cite{khomh2009exploratory,fowler2018refactoring}, including duplicated code, long methods, feature envy, large classes, message chains, and speculative generality, as summarized in Table \ref{tab:antipatterns-maintainability}.

\begin{table}[!ht]
    \centering
    \scriptsize
    \caption{Antipatterns for maintainability \cite{khomh2009exploratory,fowler2018refactoring}}
    \vspace{-2mm}
    \label{tab:antipatterns-maintainability}
    \begin{tabular}{p{11.9cm}}
    \toprule
\textbf{Mysterious Name:} A name that doesn’t clearly explain its purpose.   \\
\textbf{Duplicated Code:} The same code structure appears in multiple places. \\
\textbf{Long Function:} A function that tries to do too much and becomes hard to understand. \\ 
\textbf{Long Parameter List:} A function requires too many parameters to operate. \\ 
\textbf{Global Data:} Data accessible everywhere, making code fragile and hard to track. \\ 
\textbf{Mutable Data:} Data that changes unexpectedly, leading to bugs and side effects. \\ 
\textbf{Divergent Change:} One module often needs many modifications for different reasons. \\ 
\textbf{Shotgun Surgery:} A single change requires edits across many different classes. \\ 
\textbf{Feature Envy:} A method uses another object's data more than its own. \\ 
\textbf{Data Clumps:} Groups of data items that always appear together. \\ 
\textbf{Primitive Obsession:} Overuse of basic types instead of small, meaningful objects. \\ 
\textbf{Repeated Switches:} Multiple conditional statements checking the same conditions. \\ 
\textbf{Loops:} Manual loops that obscure intent and could be replaced with higher-level constructs. \\ 
\textbf{Lazy Element:} A class or method that no longer justifies its existence. \\ 
\textbf{Speculative Generality:} Code created just in case without real need. \\ 
\textbf{Temporary Field:} An object with fields used only in certain situations. \\ 
\textbf{Message Chains:} Navigation through multiple objects to get data. \\ 
\textbf{Middle Man:} A class that delegates all its work to others without adding value. \\ 
\textbf{Insider Trading:} Classes overly reliant on each other’s internal details. \\ 
\textbf{Large Class:} A class doing too much, becoming bloated and complex. \\ 
\textbf{Alternative Classes with Different Interfaces:} Classes similar in behavior but exposing inconsistent interfaces. \\ 
\textbf{Data Class:} A class with only fields and no meaningful behavior. \\ 
\textbf{Refused Bequest:} A subclass that inherits methods or fields it doesn’t need. \\ 
\textbf{Comments:} Comments are used to excuse unclear/messy code instead of improving it. \\ 

\bottomrule
	
	\end{tabular}
\end{table}

\noindent\paragraph{\textbf{C. Reliability Expert:}} The reliability agent evaluates robustness-related concerns, particularly exception handling or fault management practices. The assessment focuses on identifying antipatterns \cite{de2017studying}, such as over-catch of exceptions, empty handlers, generic exception propagation, and improper logging strategies. The antipattern catalog used by this agent is presented in Table \ref{tab:antipatterns-reliability}.

\begin{table}[!ht]
    \centering
    \scriptsize
    \caption{Antipatterns for reliability \cite{de2017studying}}
    \vspace{-2mm}
    \label{tab:antipatterns-reliability}
    \begin{tabular}{p{11.9cm}}
    \toprule
\textbf{Over-catch:} The handler catches multiple different lower-level exceptions. \\ 
\textbf{Over-catch and Abort:} Besides over-catching, the handler aborts the system. \\ 
\textbf{Unhandled Exceptions:} The handler does not catch all possible exceptions. \\ 
\textbf{Unreachable Handler:} The handler does not catch any possible exception. \\ 
\textbf{Catch and Do Nothing:} The handler is empty. \\ 
\textbf{Catch and Return Null:} The handler contains return null. \\ 
\textbf{Catch Generic:} The handler catches a generic exception type. \\ 
\textbf{Destructive Wrapping:} The handler propagates the exception as a new exception. \\ 
\textbf{Dummy Handler:} The handler only displays or logs some information. \\ 
\textbf{Ignoring InterruptedException:} The handler catches InterruptedException and ignores it. \\ 
\textbf{Incomplete Implementation:} The handler only contains TODO or FIXME comments. \\ 
\textbf{Log and Return Null:} Besides being a dummy handler, the handler returns null. \\ 
\textbf{Log and Throw:} The handler logs some information and propagates the exception. \\ 
\textbf{Multi-Line Log:} The handler divides log information into multiple log messages. \\ 
\textbf{Nested Try:} The handler and its try block are enclosed in another try block. \\ 
\textbf{Throw within Finally:} The handler is followed by a finally block that propagates exceptions. \\ 
\textbf{Throws Generic:} The throws propagates a generic exception type. \\ 
\textbf{Throws Kitchen Sink:} The throws propagates multiple exceptions. \\ 
\bottomrule
	\end{tabular}
\end{table}

\noindent\paragraph{\textbf{D. Performance Expert:}} The performance agent analyzes code changes from an efficiency and resource utilization perspective. The analysis targets common software performance antipatterns \cite{sun2025bitsai,wert2018performance,avritzer2022scalability,smith2012software}, including inefficient database access, excessive I/O operations, repeated calculations, unoptimized loops, inappropriate data structures, and excessive synchronization overhead. The antipatterns considered in this study are summarized in Table \ref{tab:antipatterns-performance}.

\begin{table}[!ht]
    \centering
    \scriptsize
    \caption{Antipatterns for performance \cite{sun2025bitsai,wert2018performance,avritzer2022scalability,smith2012software}}
    \vspace{-2mm}
    \label{tab:antipatterns-performance}
    \begin{tabular}{p{11.9cm}}
    \toprule
\textbf{The Stifle:} Data is retrieved by means of many similar (or equal) database queries. \\
\textbf{Expensive Database Call:} A single long-running database request causes performance overhead. \\
\textbf{Empty Semi Trucks:} An excessive number of requests is required to perform a task. \\
\textbf{The Blob:} A single class/component performs all of the work or holds all of the application’s data. \\
\textbf{Circuitous Treasure Hunt:}	A high amount of requests to retrieve the data. \\
\textbf{Wrong Cache:} Memory pollution through improper use of a cache. \\
\textbf{One Lane Bridge:} Mutual access to a shared resource is badly designed. \\
\textbf{Inappropriate Data Structures:} Using data structures that are inefficient for the required operations, causing unnecessary performance overhead. \\
\textbf{Unoptimized Loops:} Writing loops that perform excessive work or avoid optimizations, leading to slow and wasteful execution. \\
\textbf{Data Format Conversion Performance:} Repeatedly converting data between formats in costly ways that degrade runtime efficiency. \\
\textbf{Excessive or Improper Lock Usage:} Overusing or misusing locks, creating thread contention, and reducing parallel performance. \\
\textbf{Excessive I/O Operations:} Performing more disk or network I/O than necessary, resulting in significant latency and bottlenecks. \\
\textbf{Repeated Calculations:} Recomputing values unnecessarily instead of caching or reusing results, wasting CPU cycles.
\\ \bottomrule
	\end{tabular}
\end{table}

\subsection{Implementation details}
\label{sub:Implementation}
The proposed framework uses a multi-agent architecture coordinated through an LLM-based orchestration layer. The prototype is implemented using Kiro CLI\footnote{https://kiro.dev/cli/}. Kiro CLI is a terminal-based AI coding assistant that supports multi-agent workflows through declarative agent definitions, skill-based behavior specification, and sub-agent delegation. Kiro CLI provides a runtime environment where agents are defined as JSON configurations with associated skill documents (markdown-based behavioral specifications), enabling reproducible and auditable agent interactions without custom code. The system integrates with external data sources via the Model Context Protocol (MCP), an open standard for connecting LLM-based agents to tools and data services. In the following, we describe each component of our framework.

\begin{figure}[!ht]
    \centering
    \includegraphics[width=0.99\textwidth]{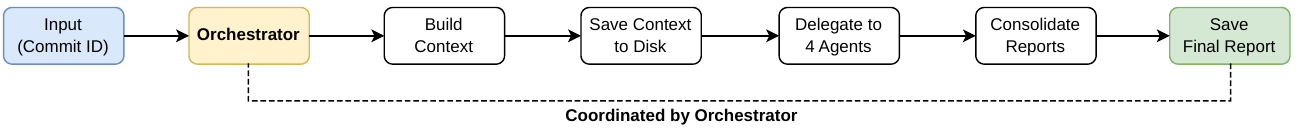}
    \vspace{-2mm}
    \caption{Code review orchestration workflow}
    \label{fig:orchestration-flow}
\end{figure}

\paragraph{\textbf{LLM orchestrator:}} The orchestrator serves as the central coordination component of the framework (see an overview in Figure \ref{fig:orchestration-flow}). It is defined declaratively through a JSON agent configuration and a persona document that specifies its behavioral rules. The orchestrator agent configuration includes the agent's skill, tools, MCP servers, and other relevant settings.
When a user prompt (e.g., review the following code commit: \#et5253e) is submitted, the orchestrator first analyzes it and determines which specialized agents should participate in the review. Then it invokes the context builder to retrieve the required contextual information and makes the relevant context available to each agent. After the specialized agents complete their analyses, the orchestrator aggregates the output into a unified review report. The report contains a summary of the review, e.g., the number of issues found per dimension, followed by the issues identified for each dimension in the four-field format shown in the first part of Table \ref{tab:feedback-template}.%following a defined template.

\paragraph{\textbf{MCP servers:}} In this study, two MCP servers are used: GitLab and Code Knowledge Graph. 

\noindent\textbf{(a) GitLab MCP server:} A containerized server (deployed via Docker) that exposes GitLab API operations as MCP tools. It provides access to the code repository, for example, to get commit metadata, code diffs, and changed files. This information provides the local context associated with a code change. 

\noindent\textbf{(b) Code Knowledge Graph MCP server:} A containerized server that exposes a code knowledge graph as an MCP tool. The knowledge graph indexes repository structures, file-level import relationships, symbol definitions (DEFINES edges), and external library usage. It supports the following operations: dependency impact analysis (fan-in/fan-out per file), folder structure exploration, AI-generated file/folder summaries, semantic/hybrid code search across repositories, and direct Cypher query execution.

\paragraph{\textbf{Context builder:}} The context builder is responsible for collecting, organizing, and preparing contextual information for agents. For a code change, the context builder retrieves the code diff, modified files, and repository metadata from GitLab. Dependency information and structural relationships are additionally retrieved using the code graph. The information collected is transformed into agent-specific contextual prompts. Different agents receive different contextual views depending on their review objectives. For example, the readability agent focuses on local code structure and formatting, whereas the maintainability agent also receive information about dependencies and structural relationships. This selective context construction aims to reduce prompt complexity while preserving the contextual information most relevant to each review dimension.

\paragraph{\textbf{Specialized review agents:}} As shown in Figure \ref{fig:core}, we implement four specialized LLM agents: Readability Expert, Maintainability Expert, Reliability Expert, and Performance Expert. Each agent operates independently and is guided through dedicated prompts, antipattern definitions, and review instructions. We use the Agent Skills\footnote{https://agentskills.io/home} protocol for all agents. Each agent is defined by a JSON configuration that specifies its name, system prompt, attached skill document (containing the antipattern checklist), and tool permissions. Agents produce findings in a strict four-field format (Antipattern Name, Location, Problem Description, Fix Suggestion). The agents are invoked as sub-agents by the orchestrator and run in parallel, each receiving only the path to the shared context document and source files on disk.

\section{Evaluation results}\label{sec:results}
In this section, we present the evaluation results of the proposed approach.

\begin{table}[!ht]
    \centering
    \scriptsize
    \caption{Identified issues per dimension from the studied  code commits}
    \vspace{-2mm}
    \label{tab:issues-count}
    \begin{tabular}{cc}
    \toprule
\textbf{Dimension} & \textbf{Count (\%)} \\ \midrule
Reliability	& 64 (31.1\%) \\
Readability	& 58 (28.2\%) \\
Maintainability	& 50 (24.3\%) \\
Performance	& 34 (16.5\%) \\ \midrule

Total & 206 \\ \bottomrule
	
	\end{tabular}
\end{table}

\subsection{RQ1: What design choices are essential for providing multifaceted and contextualized code review?}
RQ1 examines which design choices are essential for a multifaceted and contextualized code review, based on observations during solution validation.

\noindent\textbf{Orchestration and output format:} All commits were processed through the complete workflow, from agent selection to aggregation into a unified report. The orchestration is a single point of failure, as an error at any stage propagates to the final review, even when the individual agents reason correctly. Validating the context builder and the report format was therefore necessary, and providing agents with positive and negative examples was an effective way to constrain their output.

\noindent\textbf{Specialized review agents:} Each agent contributed findings (see Table~\ref{tab:issues-count}) that the other agents did not, and a review targeting a single dimension would have covered at most about a third of what was reported. Decomposing the review into agents with dedicated prompts, antipattern catalogs, and review instructions is thus what enables multifaceted assessment.

\noindent\textbf{Contextual grounding:} Only 4\% of the identified issues were judged incorrect. We attribute this low rate to the contextual information provided through the context builder and the code knowledge graph. Rather than reasoning over isolated diffs, each agent received the local context and, where relevant, structural and dependency information about the surrounding system, which appears to reduce hallucinated comments. Providing each agent with a contextual view tailored to its dimension, rather than a single undifferentiated context, also reduced prompt complexity.

\textit{These observations relate to the design presented in this study. Since neither an ablation nor a comparison with alternative designs was performed, they provide feasibility evidence rather than controlled comparative proof.}

\subsection{RQ2: How accurate is the proposed approach in identifying issues from code commits?}
RQ2 examines whether the issues reported by the proposed approach describe real concerns in the reviewed code rather than false or hallucinated findings.
 
As shown in Table~\ref{tab:correctness-validation}, 197 of the 206 issues were confirmed correct, resulting in an overall accuracy of approximately 96\%. Only 9 issues (approximately 4\%) were judged incorrect. This low rate of incorrect findings is particularly relevant in an industrial context, where reviewers tend to disengage from automated tools that produce frequent false positives. These results indicate that the approach achieves a level of accuracy that could make it viable for practical use. However, further validation across more diverse projects/contexts, and across programming languages, is needed to strengthen the generalizability of these findings.

\begin{table}[!ht]
    \centering
    \small
    \caption{Correctness of the identified issues}
    \vspace{-2mm}
    \label{tab:correctness-validation}
    \begin{tabular}{lc}
    \toprule
\textbf{Item} & \textbf{Count (\%)} \\ \midrule
Correct	& 197/206 (96\%) \\
Incorrect	& 9/206 (4\%) \\
\bottomrule
	\end{tabular}
\end{table}

\subsection{RQ3: How effective is the proposed approach in terms of identifying important issues from code commits?}
\begin{table}[!ht]
    \centering
    \small
    \caption{Importance of the correctly identified issues (197/206)}
    \vspace{-2mm}
    \label{tab:usefulness-validation}
    \begin{tabular}{lcc}
    \toprule
 \textbf{Item} & \textbf{Count (\%)} \\ \midrule
Minor --- Minor or low-impact suggestion & 62 (31\%) \\
Medium --- Important issue that should be fixed	& 70 (36\%) \\
High --- Severe issue that must be fixed & 65 (33\%) \\
\bottomrule

	\end{tabular}
\end{table}

A correct finding is not necessarily a valuable one. An automated reviewer may report accurate but trivial issues that add little value and increase the reviewer's effort. RQ3, therefore, examines the practical importance of the findings. For this analysis, we consider only the 197 issues confirmed as correct in RQ2 and examine the importance ratings assigned by the developers (reviewers) on a three-level scale: \textit{minor} (low-impact suggestion), \textit{medium} (important issue that should be fixed), and \textit{high} (severe issue that must be fixed).
 
Table~\ref{tab:usefulness-validation} presents the results for RQ3. Among the 197 correct issues, 33\% (65) were rated as high importance (issues that must be fixed) and 36\% (70) were rated as medium importance (issues that should be fixed). Together, these two categories account for approximately 69\% of the correctly identified issues, indicating that the majority of the findings were considered important by the developers and required addressing. The remaining 31\% (62) issues were rated as minor or low-impact suggestions. While minor findings may not be critical, they are not without value, as they often relate to readability and stylistic concerns that contribute to the long-term health of the codebase.

In general, the combination of high accuracy (RQ2) and a high proportion of important findings (RQ3) suggests that the approach produces reviews that are reliable and meaningful in practice. In addition, qualitative feedback from the developers corroborated these results. The reviewers were positive and highlighted the usefulness of the generated reviews, for example:
\begin{itemize}
    \item Comment-1: \textit{"The review comments were really good, especially for the maintainability and performance part."}
    \item Comment-2: \textit{"A portion of the above comments were almost the same as we had identified but your reviews had more. This was really good."}
\end{itemize}

\section{Discussion}\label{sec:discussion}
The results of this study indicate that combining specialized agent skills with context-specific knowledge yields accurate, practically relevant code reviews. The approach achieved approximately 96\% accuracy in correctly identifying issues in the investigated code commits. Additionally, among the correctly identified issues, approximately 69\% were rated as important issues (33\% severe, which must be fixed, and 36\% important with a medium rating, which should be fixed).
The low rate of incorrect findings can be attributed to the contextual grounding described in Section \ref{sec:results}. Supplying agents with repository-specific information, rather than code change alone, appears to constrain their reasoning to what actually holds in the surrounding system.
Previous work also highlights the role of contextual information in generating accurate code reviews  \cite{cihan2025automated,hu2026benchmarking,tantithamthavorn2026rovodev,ramesh2025automated}. For example, Cihan et al. \cite{cihan2025automated} investigated the impact of supplementary descriptions (e.g., comments or pull request descriptions) on the accuracy of LLM-based code review. They found that providing such descriptions improved accuracy from 59.62\% to 68.50\% for GPT-4o and from 55.56\% to 63.89\% for Gemini.

The low false positive rate is particularly important in practice, as automated reviewers that produce frequent false positives tend to erode developers' trust, increase review noise, and ultimately cause developers to disengage from such tools.
In addition, the multifaceted assessment proved valuable in identifying concerns that a single-perspective review would have missed. The identified issues were distributed on all four dimensions: reliability (31.1\%), readability (28.2\%), maintainability (24.3\%), and performance (16.5\%). This breadth more closely mirrors how human reviewers reason in practice, where multiple quality concerns are considered simultaneously rather than in isolation.

Beyond the empirical results, developing the solution yielded the following takeaways that we believe are broadly relevant to others building agentic code review systems.
\begin{itemize}
\item \textbf{Orchestration accuracy matters as much as agent design:} An error in agent selection, context construction, or aggregation propagates to the final review regardless of how well the individual agents reason.
\item \textbf{Output format is part of the design, not a presentation detail:} Constraining findings to a strict format, supported by both positive and negative examples, made a systematic assessment of 206 issues feasible.
\item \textbf{Separation of concerns should extend beyond the agents:} Giving each agent its own prompt, antipattern catalog, and contextual view reduces prompt complexity and keeps the solution extensible, as covering an additional quality dimension amounts to adding an agent rather than modifying the existing ones.
\end{itemize}

\section{Threats to validity}\label{sec:validitythreats}

\textit{\textbf{Construct validity:}} We study whether the generated reviews are trustworthy and useful to developers in practice. We operationalize trustworthiness as the correctness of each identified issue (RQ2), i.e., whether it describes a real property of the reviewed code rather than a hallucinated one, and usefulness as its importance on a three-level scale (RQ3), i.e., whether a correct finding is worth acting upon. Both measures rely on human judgment, which we consider appropriate for our research questions, since whether a finding is accurate and worth fixing depends on the semantics of the change and on system-specific knowledge.

\textit{\textbf{Internal validity:}} Having the authors (developers of the code commits) validate the findings in their own code can introduce bias in either direction, that is, greater leniency or greater scrutiny. However, the authors are best positioned to judge the relevance of a finding given their knowledge of the change.

\textit{\textbf{External validity:}} This study has several limitations that should be considered when interpreting the findings. First, the evaluation was conducted in a single company using seven commits from four projects, and the code studied was limited to Python. Although this provides realistic industrial evidence, the results may not generalize to other organizations or programming languages, and validation in other contexts is needed to assess external validity. Second, the results are based on a single LLM (Kiro CLI's default model), and the reported accuracy or importance ratings may vary with other models/frameworks. A systematic comparison across models/frameworks covering several aspects such as accuracy, ease of use, cost, and privacy would be important for organizations weighing adoption and provides a clear direction for future work. In addition, in this study, we have considered only the four dimensions of code quality. Future work may include other aspects, such as security, technical debt, and testability.

\textit{\textbf{Reliability:}} A threat specific to LLM-based solutions is non-determinism: repeated runs on the same commit may not produce identical findings. To support repeatability, the agents are specified declaratively via JSON configurations and skill documents, and the antipattern catalogs used by each agent are reported in Section \ref{sub:Review_dimensions}, enabling the setup to be reconstructed. The data collection instrument is also reported in full (Table \ref{tab:feedback-template}). However, the code commits and the generated reviews are proprietary and cannot be shared, which limits independent replication.

\section{Conclusion and future work}\label{sec:conclusion}
In this study, we proposed a contextualized multi-agent-based solution that provides a multifaceted assessment of code changes. The approach combines specialized review agents with context-specific knowledge to identify antipatterns in four dimensions: readability, maintainability, reliability, and performance. We developed and evaluated the solution in an industrial setting. The evaluation results for over 200 issues showed that the proposed solution achieves approximately 96\% accuracy in correctly identifying issues. In addition, around 69\% of the correctly identified issues were rated as important issues (with 33\% considered severe that must be fixed and 36\% important issues that should be fixed). Qualitative feedback from the developers corroborated these results and highlighted the usefulness of the generated reviews.

In future work (the next iteration of the design science research process), we plan to evaluate our solution in other contexts, including other programming languages and projects, and to include additional code quality aspects, such as security and technical debt. We also aim to implement feedback mechanisms that enable the system to learn from developers' feedback. Finally, we plan to extend the solution to support InnerSource projects, in which core teams must ensure that contributions from developers outside the project align with the project's conventions, architecture, and design decisions.

%In future work, we plan to evaluate our solution in other programming languages (e.g., Java) and to include additional code quality aspects, such as security and technical debt. Additionally, we aim to implement feedback mechanisms that enable the system to learn from developers' feedback, allowing for continuous improvement of the review recommendations over time.

%In future work, we plan to evaluate our solution in other contexts, including other programming languages and projects, and to cover additional quality aspects, e.g., security and technical debt. We also aim to implement feedback mechanisms that enable the system to learn from developers' assessments. Finally, we plan to extend the solution to support InnerSource projects, where core teams must ensure that contributions from outside developers align with the project's conventions, architecture, and design decisions.

\begin{credits}
\subsubsection{\ackname} This work was partially supported by the Knowledge Foundation through the InScale project (reference number 20230095) at Blekinge Institute of Technology (BTH), Sweden.

\subsubsection{\discintname}
The authors declare that they have no known competing financial interests or personal relationships that could have influenced the work reported in this paper.
\end{credits}

\bibliographystyle{splncs04}
\bibliography{paper}
\end{document}